\documentclass[12pt]{article}
\usepackage{epsfig}
\begin{document}

\setlength{\baselineskip}{0.30in}
\def\psl{p \hspace*{-0.5em}/}
\def\ksl{k \hspace*{-0.5em}/}
\newcommand{\be}{\begin{equation}}
\newcommand{\bi}{\bibitem}
\newcommand{\ee}{\end{equation}}
\newcommand{\ra}{\rightarrow}

\begin{flushright}
UM-TH-00-14 \\ 
Freiburg-THEP-00/15.\\
\end{flushright}

\begin{center}
\vglue .06in
{\Large \bf {Interplay between Perturbative and Non-perturbative Effects in
the Stealthy Higgs Model}}\\ [.5in]

{\bf R. Akhoury$^{(1)}$, J.J. van der Bij$^{(1,2)}$
 and H. Wang$^{(1)}$}\\ [.15in]

{\it $^{(1)}$ The Randall Laboratory of Physics\\
University of Michigan\\
Ann Arbor, MI 48109-1120}\\ [.15in]
{\it and}\\ [.15in]
{\it $^{(2)}$ Fakult\"at f\"ur Physik\\
Universit\"at Freiburg\\
H. Herderstr. 3, 79104 Freiburg\\}
\end{center}
\begin{abstract}
\begin{quotation}
We study corrections to electroweak precision variables
in a model with strongly interacting singlet Higgs particles. 
\end{quotation}
\end{abstract}
\newpage

\section{Introduction }
With the close of the LEP experiments the structure of the weak
interactions as a gauge theory has been fully confirmed. The only 
missing ingredient at the moment is direct evidence of the Higgs 
sector. The LEP-200 experiment gives a lower bound of $m_H > 113
GeV$\cite{prec1}. The precision measurements at LEP-1 and SLC imply a
relatively low Higgs mass ( $< 170 GeV)$\cite{prec2},
 though some lingering doubts remain because of the
different values for the hadronic and leptonic data.

If the Higgs boson is indeed light and of the standard model type,
it should be easy to find at the LHC. However the precision data
do not prove that this is indeed the case. There are still two
possibilities to hide the Higgs boson at the LHC. The first is that
the Higgs boson is simply too heavy to be produced. Within the standard
model this implies strong interactions. In that case the limit
on the Higgs boson mass, which comes from one-loop calculations
is not reliable. While the two-loop corrections are small and cannot
provide for a fit to a heavy Higgs-boson, the full higher order
results could be different. There are some indications in the literature
that resumming bubble-graphs in the Higgs propagator can lead to
a saturation of the radiative corrections. The second way to hide
a Higgs boson at the LHC is by the introduction of singlet Higgs fields
ref.[3-11].
In this case there are two effects. One is mixing of the doublet and 
singlet Higgs boson, leading to a split of the Higgs peak into different
peaks, each being less significant than a single standard model peak.
The second effect is the possibility of decay into invisible singlet
particles. The invisible decay width is not necessarily small, so
 one could have a light Higgs particle with a large invisible width.
One can also combine mixing and invisible decay, to generate a Higgs
signal spread out over an arbitrary energy range with a large invisible 
decay fraction. Such a Higgs signal would be extremely hard to identify
at the LHC, since there is no peak  in the signal compared to the 
background. The signal would be an overall enhancement of missing energy.
For this to be useful, one has to know the background very precisely.
The background cannot be calculated precisely at the LHC; however at a high
energy $e^+ e^-$ collider there is no problem to look for this signal. It
is important to point out here that an additional phenomenological
advantage of introducing the singlet Higgs fields is that they would be
prime candidates for self-interacting cold dark matter. In fact, the
model introduced in the next section is the simplest one with WIMPS.

There is therefore the realistic possibility that the LHC will not
see  evidence for the Higgs sector. That means that the only information
would consist of the LEP precision measurements. It is for this reason,
that we decided to look more carefully at the radiative corrections
in the invisible Higgs scenario. As both a large width and strong 
interactions can play a role we decided to study the radiative corrections
in the so-called stealthy Higgs model \cite{binoth}. 

We describe the model in chapter 2. In chapter 3 we discuss some
analytical results for the simplest case, to gain some understanding
on the difference between the all order bubble resummation from the
1/N expansion and the results of finite order in perturbation theory.
In chapter 4 we give numerical results. 
In chapter 5 we discuss the results and give our conclusions.

\section{The Higgs--$O(N)$-Singlet Model} 
To illustrate the consequences of a hidden sector coupled to the Higgs boson
in a possibly strong way, we want to consider the case of scalar gauge singlets
-- let us call them ''Phions'', -- added to the SM. 
To deal with the  case of strong interactions
we introduce an $N$--plet
of such Phions. This allows us to use nonperturbative $1/N$--methods.
Neglecting all the fermions and gauge couplings for the moment, 
our model consists  of the
SM Higgs sector coupled to an $O(N)$--symmetric scalar model. 
Similar models can be found in Ref.~\cite{chiv1,chiv2,bjorken}.
Our Lagrangian density is:
\begin{eqnarray} \label{model}
L_{Scalar} &=& L_{Higgs} + L_{Phion} + L_{Interaction}   \nonumber  \\
           & & \hspace{7cm} \mbox{where}\nonumber \\
L_{Higgs}  &=& 
 - \partial_{\mu}\phi^+ \partial^{\mu}\phi -\lambda \,
 (\phi^+\phi - \frac{v^2}{2})^2 \nonumber \\
L_{Phion}  &=& - \frac{1}{2}\,\partial_{\mu} \vec\varphi \, 
\partial^{\mu}\vec\varphi
     -\frac{1}{2} m_{P}^2 \,\vec\varphi^2 - \frac{\kappa}{8N} \, 
     (\vec\varphi^2 )^2 \nonumber \\
L_{Inter.} &=& -\frac{\omega}{2\sqrt{N}}\, \, \vec\varphi^2 \,\phi^+\phi 
\end{eqnarray}  
Here we use a metric with signature $(-+++)$. 
$\phi=(\sigma+v+i\pi_1,\pi_2+i\pi_3)/\sqrt{2}$ is the complex Higgs doublet of 
the SM with the
vacuum expectation value $<0|\phi|0> = (v/\sqrt{2},0)$, $v=246$ GeV. Here, 
$\sigma$ is the physical  
Higgs boson and $\pi_{i=1,2,3}$ are the three Goldstone bosons. 
$\vec\varphi = (\varphi_1,\dots,\varphi_N)$ is a real vector with 
$<0|\vec\varphi|0>= \vec 0$. If we would allow for a non-vanishing vacuum expectation 
value for the Phions, the mass matrix
would become non-diagonal and Higgs--Phion mixings would occur. 
The lightest scalar of the gauged model
would have a reduced coupling to the vector bosons by a 
cosine of a mixing angle. We will not discuss this possibility 
further, as we are mainly interested in the effects coming from the Higgs width.
If we look at the gauged model we can choose the unitary gauge to rotate away 
the unphysical Goldstone bosons. This is gauge invariant, because in the 
following we only consider loops
of gauge singlet particles. Note that the vacuum induced mass term for the 
Phions is suppressed by a factor $1/\sqrt{N}$.

In the case of large non standard couplings $\omega$ and $\kappa$, 
loop induced operators with external Higgs and Phion fields appear and are not 
negligible. They are only suppressed by powers of $1/N$. 
For the discussion of Higgs signatures, it is enough to focus on the 
Higgs-propagator.

As shown above the propagator is modified by the Phions.
In the leading order in $1/N$, which is found in the limit $N \rightarrow 
\infty$, the Higgs self-energy is given by an infinite sum of Phion bubble 
terms. Regularization of the divergent bubbles, i. e. absorbing the divergent 
and some constant
contributions into the bare parameters, is done by subtraction of the 
logarithmically divergent part  . 
With this regularization, the Euclidean bubble integral
\begin{displaymath} 
       I_{Bubble}(s=-p^2,m_{\phi}^2) = \frac{1}{2} \int \frac{d^4k}{(2\pi)^4}\,
                 \frac{1}{k^2+m_{\phi}^2}\,\frac{1}{(k+p)^2+m_{\phi}^2}  \nonumber,
\end{displaymath}
becomes above the Phion threshold
\begin{eqnarray}
 && I(s,\mu^2,m_{\phi}^2) \quad = \quad  I_{Bubble}(s,m_{\phi}^2) -
I_{Bubble}(0,\mu^2)\\
    \qquad           && = -\frac{1}{32\pi^2}  \left(
\log(\frac{m_{\phi}^2}{\mu^2}) 
    - 2 +  \sqrt{1-\frac{4m_{\phi}^2}{s}} 
           \left( \log \left(          
\frac{1+\sqrt{1-\frac{4m_{\phi}^2}{s}}}{1-\sqrt{1-\frac{4m_{\phi}^2}{s}}}
           \right)  - i\pi\right)\right)\nonumber 
\end{eqnarray}
with the arbitrary renormalization scale $\mu$. 
In the case of massless Phions this simply reduces to  
$I(s,\mu^2,0) = -1/(32\pi^2)(\log(s/(e\mu)^2) - i\pi)$. The bubble sum is 
the geometric series of the integral times a coupling.\\
Adding up all regularized terms gives the inverse Higgs propagator     
\begin{eqnarray} \label{propagator}
D_{H}^{-1}(s,\mu^2) &=& -s + M_H^2 - i\sqrt{s}\Gamma_{SM}(s)
                     + \Sigma (s,\mu^2)  \\
                     && \nonumber \\
\Sigma (s,\mu^2) &=&
 \frac{-\omega^2 v^2 I(s,\mu^2,m_{\phi}^2)}{1+\kappa I(s,\mu^2,m_{\phi}^2)}
\nonumber
\end{eqnarray}
Above the Phion threshold, $s>4\,m_{\phi}^2$, $\Sigma$
develops an imaginary part which results in a Higgs width depending on the non 
standard parameters leading to observable effects.
The independent SM Higgs-width is added, too.
To find an explicit expression for  the upper propagator, remember that within 
the SM the Higgs mass, or better the quartic Higgs coupling,
is a free parameter. 

Defining the mass by the location of the resonance on the real $p^2$--axis 
fixes our renormalization scale $\mu$ by the equation
\begin{equation}\label{rencon}
Re( \Sigma (M_H^2,\mu^2) ) = 0
\end{equation}
Using this relation, the abbreviations  
$\tilde\omega^2=\omega^2/(32\pi^2)$, 
$\tilde\kappa=\kappa/(32\pi^2)$ and $r(x) = \sqrt{1 - 4\,m_{\phi}^2/x}$,
one finds, after splitting the integral in its real and imaginary part
\begin{eqnarray} 
I(s,\mu^2,m_{\phi}^2)\vert_{\mu\,fixed} &=& a(s)+i\,b(s) \nonumber\\
a(s) &=&  (\sqrt{1-(2\pi\tilde\kappa r(M_H^2))^2}-1)/(2\tilde\kappa) \nonumber\\
  &&  + r(M_H^2) \log(\frac{1+r(M_H^2)}{1-r(M_H^2)})- r(s) \log(\frac{1+r(s)}{1-r(s)})\nonumber\\
b(s) &=&  \pi\,r(s)  \, , \nonumber
\end{eqnarray}  
an expression for the Higgs propagator, in terms 
of running quantities:
\begin{eqnarray} 
D_{H}^{-1}(s) &=& -s + M_H(s)^2 - i\sqrt{s}\,\Gamma_{H}(s) \\ \nonumber\\
M_H(s)^2 &=& M_H^2 - \tilde\omega^2v^2  
\frac{a(s) + \tilde\kappa (a(s)^2+b(s)^2)}{(1+\tilde\kappa\,a(s))^2+
(\tilde\kappa\,b(s))^2} \nonumber\\
\Gamma_{H}(s) &=& \Gamma_{SM}(s)+\frac{\tilde\omega^2v^2}{\sqrt{s}}
\frac{b(s)}{(1+\tilde\kappa a(s))^2+(\tilde\kappa\,b(s))^2}\nonumber
\end{eqnarray}  
Remember that this expression is only valid above the Phion threshold.

The advantage of this model is that one can study separately the effect
of a strong coupling $\omega$ of the standard model Higgs to the hidden
 sector and of strong interactions $\kappa$ within the hidden sector. 
Thereby one separates the effects of a large width from the effects
of strong interactions.
We mention that the standard model in the 1/N expansion \cite{ein1,ein2,cho,aoki}
is reproduced within this model by taking $m_H^2 = 2 \lambda v^2$ 
and $ \omega = \kappa = 2 \lambda$.

\section{Analytical Considerations.}
In this chapter we discuss some analytical results.
At first sight this would appear to be a hopeless undertaking,
because of the complicated form of the propagator. Also there 
seems to be no reason to do so. Naively one would take simply 
the non-perturbative Higgs-propagator, insert it in the diagrams
and calculate the result numerically. This has indeed been tried
in the literature \cite{ein2,cho, aoki}.
 However this procedure has a problem due to
the presence of a tachyon in the propagator and  one 
will not get finite results. In order to better understand what is
going on, it is therefore advantageous to attempt an analytic calculation.
This is clearly not possible in the general case. We will study therefore
the simplified case $m_{P} = 0$ and $\kappa =0$. Physically we
have an unstable Higgs-particle with a width determined by the 
coupling $\omega$. The Higgs-propagator simplifies to:
$$ D_H^{-1}= -s +m_H^2 +\Gamma  log(- s/m^2_H -i \epsilon) $$

with  $\Gamma= \frac{\omega^2 v^2}{32 \pi^2}$.
We limit the discussion in this chapter to the so-called $\rho$
parameter, which is the ratio of neutral to charged current strengths.
$$\rho = G_F^0 / G_F^+$$
At the tree-level $\rho = 1$; the correction is given by
$$\delta \rho = \rho -1 = (\delta M_W^2 - cos^2(\theta_w) \delta 
M_Z^2) / M_W^2 $$
where $\delta M_W^2$ and $\delta M_Z^2$ are the corrections to the
vectorboson masses ate $k^2=0$.
The $\rho$-parameter is one of the parameters measured in the electroweak tests. It is related to the commonly used T-parameter by
$T = (1-\rho^{-1}) / \alpha$.

\subsection{The Standard Model}
Within the standard  model the one-loop Higgs mass dependent 
correction to $\delta \rho$ is simplest decribed in the unitary 
gauge. Here only one diagram contributes, the tadpole like graphs
containing the $2H-2W$ -vertex cancel in $\delta \rho$.
One finds, using dimensional regularization, with n the dimension
of spacetime :
$$ \delta \rho =\frac {-g^2}{(2\pi)^4i} (1-\frac{1}{n}) (\int d^nk (k^2+M_W^2)^{-1}(k^2+m_H^2)^{-1} 
-\frac{1}{cos^2(\theta_W)} \int d^nk (k^2+M_Z^2)^{-1}(k^2+m_H^2)^{-1}) 
$$
 The contribution from these Higgs-dependent graphs is still 
infinite, of the form:
\be
\delta \rho = -\frac {3g^2}{64 \pi^2} tg^2(\theta_W)
( 2/(n-4) + log(m_H^2) +finite )
\ee
The infinite piece cancels against the infinities coming from 
the pure W-boson graphs, that are independent of the form
of the Higgs-propagator. The explicit form of the finite part
is quite complicated. To get a simple result for the Higgs-mass 
dependence only, we subtract the contribution for the fictitious case
$m_H =0$:
$$\delta \rho (S.M.; m_H) - \delta \rho (S.M.; m_H =0) =
- \frac{3 g^2}{64 \pi^2} ( 
\frac{1}{cos^2(\theta_W)} \frac{m_H^2}{m_H^2 - M_Z^2} log(m_H^2/M_Z^2)
- \frac{m_H^2}{m_H^2 - M_W^2} log(m_H^2/M_W^2))$$

In the limit $tg(\theta_W) \rightarrow 0$ this simplifies to :
$$\delta \rho (S.M.; m_H) - \delta \rho (S.M.; m_H =0) =
- \frac{3 g^2}{64 \pi^2} tg^2(\theta_W)( \frac { m_H^2}{m_H^2 - M_W^2}
log(m_H^2/M_W^2))$$

showing the close connection between the $\rho$ parameter and the hypercharge 
coupling. If we further assume that $m_H >> M_W$ one can simply express 
the $\rho$ parameter by the integral:

$$ \delta \rho =\frac {g^2}{(2\pi)^4i} tg^2(\theta_W)(1-\frac{1}{n})\int d^nk 
(k^2)^{-1}(k^2+m_H^2)^{-1} $$
This form is useful for the more elaborate calculations later on.
In the large Higgs mass limit one finds therefore:
\be
\delta \rho (S.M.; m_H) - \delta \rho (S.M.; m_H =0) =
- \frac{3 g^2}{64 \pi^2} tg^2(\theta_W) log(m_H^2/M_W^2).
\label{deltarho1}
\ee

\subsection{Two-Loop Result}
The correction $\delta \rho_2$ at the two-loop level can be straightforwardly calculated 
using the techniques of ref.~\cite{bij1,bij2}.
 Experience from the standard model
where it was found that $\delta \rho_2 \approx m_H^2$ would lead one to
expect that in our model $\delta \rho_2 \approx \omega^2 v^2/ m_H^2$.
However an explicit calculation gives :
$$ \delta \rho_2 = \frac {3g^2 \omega^2 v^2}{4096 \pi^4} (
\frac {M_W^2} {(m_H^2-M_W^2)^2} log^2(m_H^2/M_W^2)
- \frac{1}{cos^2(\theta_W)}\frac {M_Z^2} {(m_H^2-M_Z^2)^2} log^2(m_H^2/M_Z^2) )
$$ This shows that for large Higgs mass the two-loop correction is 
suppressed compared to naive expectations. This might be a clue as to why
the standard model coefficient of the two-loop corrections to electroweak 
quantities is very small. Indeed the inclusion of the two-loop
heavy Higgs corrections within the standard model
do not significantly effect the electroweak
fits. This might therefore indicate that the first large corrections
would appear at the three-loop level in the $\rho$ parameter.
Indeed it is known from calculations 
in the standard model, that only at the two-loop level there are 
large changes in the Higgs-propagator. Since the $\rho$-parameter
probes the Higgs propagator indirectly, two-loop corrections in
the Higgs-propagator translate into thee-loop corrections in the 
$\rho$-parameter. In any case the above result shows that it is important
to check what happens at higher order.

\subsection{All Orders Perturbation Theory}
To calculate the all orders results we work in the limit where
the Higgs mass is large and $tg(\theta_W)$ small,
 so that we can ignore the mass of the
vectorbosons within the diagrams. 
Going to Euclidean space a  diagram with n phion-bubbles
can then be written in the form :
$$\int d^4k \frac{1}{k^2} \frac{1}{(k^2+m_H^2)^{n+1}} log^n(k^2/m_H^2) $$
Going to polar coordinates the $1/k^2$ factor cancels in the
$d^4k$ integration, thereby simplifying the integrals.
Keeping the coupling constants we find:
$$\delta \rho = \frac {3 g^2 tg^2(\theta_W)}{64 \pi^2}
\sum_{n=1}^{\infty} \int_{0}^{\infty} (-\Delta)^n 
 \frac {log^n(s)} {(s+1)^{n+1}} ds $$
where:
$$\Delta = \frac {\omega^2 v^2}{32 \pi^2 m_H^2} $$
The integrals can be performed explicitely to give:

$$\delta \rho = 
 \frac {3 g^2 tg^2(\theta_W)}{64 \pi^2}
\sum_{n=2}^{\infty}  \frac {(-2\pi \Delta)^n} {n} 
|B_n| ( 1 - 2^{1-n})$$
where $B_n$ are the Bernouilli numbers.
Since the odd Bernouilli numbers are 0, only the graphs with
an even number of phion-bubbles contribute, which explains the
suppression found in the previous section.
The series found above is clearly divergent. The question is
whether it is resummable. The Borel sum is defined as follows.
Given the above series $\sum_{n=2}^{\infty} a_n \Delta^n$
we form the new "Borel" series $F(z) = \sum_{n=2}^{\infty} 
a_n z^{n-1} / (n-1)! $.
We find :
$$F(z)= \frac {\pi }{sin (\pi z)} - \frac{1}{z}$$
The Borel transform has an infinity of poles for positive
values of the coupling constant $\Delta$. This means that there is
no unambiguous way to resum the perturbative series, so that
non-perturbative  effects much be present. We will return to 
the significance of this result for vacuum instability in the 
next section.

\subsection{Non-perturbative Contribution.}

Given the fact that the perturbation theory does not converge,
we try to calculate the corection to $\delta \rho$ by first
resumming the phion-bubbles within the propgator and then 
inserting the dressed propagator in the diagram. The most efficient way
to calculate $\delta \rho$ is to use the Kall\'{e}n-Lehmann
representation for the Higgs propagator:
$$D_H(k^2)= \int ds' \sigma (s')/(k^2 + s' -i\epsilon) $$
Figuratively  speaking we thereby write the Higgs propagator as the
sum (integral) of a number of Higgses with different masses.
The contribution to $\delta \rho$ is then a weighted sum over 
the different Higgs masses:
\be
\delta \rho = \int ds' \sigma(s') \delta \rho( m_H^2 = s')
\ee
where $\delta \rho(m_H^2= s')$ is given in eq.(\ref{deltarho1}).
For the propagator to be physical it is necessary that the spectral
weight  $\sigma(s')$ is positive and is unequal to zero, only for
$s' > 0$. The latter is not the case; the resummed propagator contains
at least a tachyon. Suppose that the location of a tachyon pole is given
by $m_T^2 = - s_0 m_H^2$, with $s_0$ the solution of the equation
$ s_0 + 1 + \Delta log(s_0) = 0 $. The residue at the pole is 
given by $ -s_0/(\Delta + s_0) $. Let us assume for the moment that
there is only a single tachyon pole. The pole structure of the propagator is
examined in detail in the appendix for different values of the parameters.
There it is found that depending on the parameters there can be, in general
more than one unphysical pole, however for the case at hand, i.e.,
$\tilde{\kappa}=0$  and massless Phions, there is indeed only one tachyonic 
pole.The simplest way to treat this tachyon pole is to subtract it from the
propagator. In that case one finds an, at first sight acceptable, spectral
density:
\be
\sigma(s') = \frac {\pi \Gamma}{|-s+m_H^2 +\Gamma log(s/m_H^2) -i \pi
\Gamma|^2}.
\ee
When one tries to calculate the $\rho$-parameter with this
spectral density one runs into a problem. The contribution to $\delta \rho$
coming from a single Higgs boson graph contains a divergence
$1/(n-4)$, that gets canceled by the pure vectorboson graphs.
For the generalized propagator this translates into a contribution
$1/(n-4) \int \sigma(s') ds'$. So in order to get a finite
contribution to $\delta \rho$ one needs to fulfill the condition
$\int \sigma(s') ds'= 1$. This condition is automatically fulfilled
at each order in perturbation theory, because of the renormalizability
of the theory. The condition is however not fulfilled for the
tachyon-subtracted resummed propagator. By a simple contour integral
one sees that the difference is indeed due to the tachyon pole:
\be
\int ds'  \sigma(s') = \Delta / (\Delta + s_0).
\ee
A graph of the factor as a function of $\Delta$ is given in
fig.[1].
The effect is non-perturbative as $s_0 \approx exp(-1/\Delta)$ for
$\Delta \rightarrow 0$. Also for very large width the effect 
becomes small as then $s_0 \rightarrow 1$. The effect is numerically
largest for $\Delta =1 $. 
The presence of the tachyon can be understood as an artefact 
of the approximation we made. We took into account only the 
bubble graphs connecting a Higgs-boson with two phions. This means
that we are effectively dealing with a $\phi^3$ theory, thereby having
no lowest energy state. The presence of the non-perturbative tachyon
thus signifies the possibility of the vacuum being unstable. This vacuum
instability was already indicated in the previous subsection by the
singularities of the Borel transform on the positive real axis. It is 
similar to the instability of the QED vacuum against formation of
electron-positron pairs in the presence of strong electric fields
\cite{schwinger}. Indeed such considerations already exist in the
literature \cite{olesen}. Using these results and deforming the relevant
contours it is easy to determine the imaginary part of the vacuum to vacuum
phase ($\delta$) which is related to the probability of vacuum decay. We find,
$$Im \delta~=~{-i\pi \over e^{1/\Delta}+1}.$$ This expression exhibits the same 
qualitative behaviour outlined above.

In the full
theory, the spectral density does not just contain the two-phion
cut, there are also two-Higgs and multi-phion cuts. Since the full
theory has a vacuum the tachyon-pole should disappear when all graphs
are taken into account. However finding the exact propagator would mean
solving the theory completely. This is not possible at present. As long
as we are limited to summing partial sets of graphs such instabilities 
are bound to be present. To still get a reasonable idea of the possible
effects of a large Higgs width, we have to find a phenomenological
prescription to deal with this problem. The prescription has to satisfy
two conditions, first it should reproduce perturbation theory, second
it should satisfy the Kall\'{e}n-Lehmann representation for the Higgs
propagator. We chose therefore to do the following. We start with
the resummed propagator and subtract the tachyon pole. The resulting
spectral density is positive definite and non-zero only for $s' >0$.
It is however not correctly normalized, so we multiply the spectral
density with the nonperturbative correction factor $ (\Delta +s_0)/\Delta$.
This way we keep the shape of the spectral density the same, basically
assuming that the spectral density is dominated by the two-phion states.
This is not a perfect procedure of course, but lacking the means to 
solve the theory exactly, it appears the best one can do. From the numbers
in fig.[1] we expect the result not to be too far from the truth. 

\section{Numerical results.}
The above considerations are applied in this section to the calculation of $\delta
\rho$ and of the $S$ parameter in the stealthy Higgs model.

  From the discussion in sections 2,3 we may write down
the following formulae for these parameters. Consider first the difference of the
$\rho$ parameter in the Stealthy Higgs model and in the Standard model:
\begin{eqnarray}
\delta \rho(sth,m_H) - \delta \rho(SM,m_H)~=~{3g^2 \over 64\pi^2}\int_0^{\infty}ds
\bar{\sigma}(s) \nonumber \\
\left( f(s,M^2) - {1 \over c^2}f(s,M_z^2) - f(m_H,M^2) +
{1 \over c^2}f(m_H,M_z^2) \right) \\
f(s,M^2)~=~{s \over s-M^2}\ln({s \over M^2})
\end{eqnarray}
For the case of massless phions and with $\kappa=0$, the density $\bar{\sigma}$ is,
\begin{equation}
\bar{\sigma}(s)~=~{\Gamma + s_0m_H^2 \over (-s+m_H^2+\Gamma\ln{s \over m_H^2})^2 +
\pi^2\Gamma^2}
\end{equation}
The more general cases are discussed in the appendix. There can be more than
one unphysical pole and a subtraction has to be made for all of them.The
renormalization factor was determined numerically using the expression for the
spectral density that is obtained after subtracting the tachyon pole and checked
against the expression eq.(\ref{multipole}) in the appendix.

 Customarily, the
$S$ parameter is defined by:
\begin{equation}
{\alpha \over 4c^2s^2}S~=~{\Pi_{ZZ}(M_Z^2) - \Pi_{ZZ}(0) \over M_Z^2}
\end{equation}
>From this it is straightforward to obtain the following formula for the difference of
this parameter in the stealthy Higgs model and in the SM:
\begin{eqnarray}
S(sth, m_H) - S(SM, m_H)~=~{- 1 \over \pi}\int \bar{\sigma}(s)ds (H(s) -
H(m_H^2)) \\ H(m_H^2)~=~({m_H^2 \over M_Z^2})^2 \left( -1/12-1/12 F(m_H) -{1
\over 12}\ln{M_Z^2 \over m_H^2} \right) + {m_H^2 \over m_H^2 -
M_Z^2}(-3/4\ln{M_Z^2 \over m_H^2}) \nonumber \\ + { m_H^2 \over M_Z^2} \left(
7/24 + 1/3 F(m_H) + 1/4 \ln{M_Z^2 \over m_H^2} \right) -2 - F(m_H). \\
F(m_H)~=~ \int_0^1 \ln(x^2+{m_H^2 \over M_Z^2}(1-x))
\end{eqnarray}

These formulae can now be used to study the radiative corrections
for physically interesting cases. As a first example we take
$m_H=150 GeV,~~ m_{\phi}=60 GeV,~~ \kappa=0$.
These parameters are of interest, since they correspond to a typical
case, where no evidence would have been seen at LEP, where the LHC
is insensitive, but where a linear $e^+ e^-$-collider would discover
the Higgs boson. With respect to the radiative corrections there are
two questions to be discussed here. The first is whether there are
significant differences between the Stealth model and the standard 
model. We see from figs.[2,3] that the difference is actually quite
small, of the order of a few $10^{-2}$, which is within the errors
of the measurements. The corrections behave as if 
one had a somewhat heavier Higgs as in the standard model. If a large width
$\Gamma_H$ gives a contribution, precisely mimicking a heavier Higgs, there is
of course still no way to determine, whether the precision measurements
prefer the Standard model or the Stealth model. Therefore it is 
useful to consider the Higgs mass (large Higgs mass) independent 
contribution $ 6 \pi S + 8/3 \pi cos^2(\theta_W)T$ and see if deviations
are present. This combination is given in fig.[4]. We see that a
difference is present, however it is quite small. This example
shows, that the precision tests  cannot rule out the Stealth model
as the differences between it and the standard model are small.
However the Stealth model appears to always mimick a heavier Higgs
than the Standard model Higgs, for the same value of the mass parameter.
Therefore the Stealth model cannot generate corrections that would 
behave as if the Higgs were very light, which appears preferred by 
the leptonic  data.

Another case of special interest is the Standard model in the
$1/N$ limit, which has also been discussed in \cite{ein1,ein2,cho,aoki}.
This model corresponds to $m_H^2=2 \lambda v^2, \omega=\kappa=2 \lambda$.
The graphs for the radiative corrections are  given in figs.[5-6]. It is to be
noted here that in the analysis of the $1/N$ limit, we are keeping the vector 
boson mass to be fixed. Thus in the figures for this case, we have used 
$v/\sqrt{N}$ for the vacuum expectation value.  We see that for a large range of
the Higgs mass, up to about
$1 TeV$ there is essentially no change compared to the standard model
one-loop corrections. After this scale the strong interactions take over
and have the effect, both in S and T, of increasing the large Higgs-mass
growth of the radiative corrections. This does not happen with S and T in the 
same way, so that there is a Higgs mass dependence in the sum
$ 6 \pi S + 8/3 \pi cos^2(\theta_W)T$ as seen in fig.[7]. This appears contrary to
the statements in the literature, where  a saturation of the radiative corrections
is found. There are two important points worth mentioning here. First, we
restrict ourselves in this analysis to values of approximately $m_H \leq 1$ TeV.
This because of the way we have set up the formalism in section 2- this constraint
is the only way to satisfy eq.(\ref{rencon}). It is therefore possible that the
saturation sets in at larger values of the Higgs mass. This possibility will
be studied in a separate publication. Secondly, it is not clear how previous 
authors have treated problems with the tachyon. Based on fig[7], we
therefore conclude that after resummation of the bubble-graphs, the large
Higgs-mass case appears to be ruled out by the LEP precision data. 

Finally figs.[8-10] illustrate the $\kappa$ dependence of the difference of the 
$S$ and $T$ parameters for the Stealth and Standard models. The conclusions here
are the similar to the $\kappa=0$ case.

\section{Acknowledgements}
This work was supported by the NATO grant CRG 970113, by a University of Michigan
Rackham grant to promote international partnerships,  by the US Department of
Energy and by the European Union contract HPRN-CT-2000-00149.

\newpage

\section{Appendix}
In this appendix we will discuss the question of the
renormalization of the Higgs propagator spectral density in some more detail
than in the main text.
 
It was argued in sec. 3.4 
that a necessary condition for the finiteness of the theory is :
\begin{equation}
\int \sigma(s')ds' ~=~1.
\end{equation}
Since the resummed propagator contains a tachyon pole which is an artifact of the
approximation procedure, we choose to subtract this pole. The resulting spectral
density $\sigma'(s')$ is however not properly normalized. Indeed, let us note that we 
may write for the inverse propagator near a zero at $k^2= -m^2$:
\begin{equation}
D_H^{-1}(k^2)~=~(k^2+m^2)(1+\int{1 \over k^2+\mu^2-i\epsilon}\lambda(\mu^2)d\mu^2)
\end{equation}
One can relate the $\lambda(\mu^2)$ to $\sigma(\mu^2)$ by considering the imaginary
part of the above:
\begin{equation}
Im D_H^{-1}(k^2)~=~-{Im D_H(k^2) \over |D_H(k^2)|^2}~=~{-\pi \sigma(-k^2) \over
|D_H(k^2)|^2}~=~\pi (k^2+m^2)\lambda(-k^2)
\end{equation}
or,
\begin{equation}
\lambda(\mu^2)~=~{ \sigma(\mu^2) \over (\mu^2-m^2)|D_H(k^2)|^2}
\end{equation}
The contribution to $\lambda$ from the two body cuts can now be easily obtained from
that of $\sigma$:
\begin{equation}
\lambda_2(\mu^2)~=~{\Gamma(\mu^2) \over \mu^2-m^2}
\end{equation}
In the above, for the case of massless phions with only 3 point interactions,
$\Gamma(\mu^2)=\Gamma$ introduced earlier, and for the case of massive phions, 
\begin{equation}
\Gamma(\mu^2)~=~\Gamma (1-{4m_{\phi}^2 \over \mu^2})^{1/2}
\end{equation}
We thus get the following in the approximation of keeping only the two body cuts :
\begin{equation}
D_H^{-1}(k^2)~=~(k^2+m^2)(1+\int_c^{\infty}{\Gamma(\mu^2) \over 
(k^2+\mu^2-i\epsilon)(\mu^2-m^2)} d\mu^2
\end{equation}
For massless  phions the lower cut $c$ is 0 and for massive ones it is
$4m_{\phi}^2$. The residue $z$ of the propagator at a pole $k^2=-m^2$ is given
by:($k^2=-s$)
\begin{equation}
z^{-1}~=~{- d \over ds}D_H^{-1}(s)|_{s=m^2}
\end{equation}
As an example, the residue at the tachyon pole $m_T^2=-s_0 m_H^2$ is easily seen to be
the following for massless phions:
\begin{equation}
z^{-1}~=~1 + {\Gamma \over s_0m_H^2}.
\end{equation}

Now we are ready to determine the renormalization of $\sigma'$ (call it
$\bar{\sigma}$) such that $\int\bar{\sigma}(s')ds' = 1$. Noting that at a pole
$s'=m^2$, $\sigma(s') = z\delta(s-m^2)$ we get, if such is pole is to be removed:
\begin{equation}
1~=~\int ds' \sigma(s')~=~z + \int ds' \sigma'(s').
\end{equation}
From this we see that:
\begin{equation}
\bar{\sigma}~=~ ({1 \over 1-z}) \sigma' 
\end{equation}
As discussed in section 3.4 it gives the renormalization factor of
$(\Delta+s_0)/\Delta)$ for the tachyon in the massless phion case.

Such a procedure is quite general and may be used for any number of consistent
subtractions. Thus if there are multiple unphysical poles that we wish to remove 
then the corresponding renormalized spectral density may be written as: 
\begin{equation}
\bar{\sigma}~=~ ({1 \over 1-\Sigma z_i}) \sigma' ,
\end{equation}
where, the $z_i$ denote the z-factors at the positions of these unphysical poles.

When the complete model is treated, i.e., the four point couplings are included 
and  the Phions are massive, the unphysical pole structure is much more
complicated. The  corresponding pole positions, spectral functions and the $z_i$
functions may be obtained from the description of the model in section 1. we will
summarize the  various cases below:

case (1) $\tilde{\kappa}=0, m_{\phi}=0$. In this case as discussed earlier there is
only  one tachyon pole.
case (2) $\tilde{\kappa}=0, m_{\phi} \neq 0$. Here again there is only one pole. 
Define
$$ \Gamma_2={m_H^2 \over c_H-2}$$ and $$c_H=r(m_H^2)\ln{1+r(m_H^2) \over
1-r(m_H^2)}.$$ Then when, $\Gamma >\Gamma_2>0$, the pole is a tachyon, 
otherwise when $\Gamma_2<0$ the pole is physical as well as the case when 
$\Gamma <\Gamma_2$ and $\Gamma_2 >0$.
case (3) $\tilde{\kappa} \neq 0, m_{\phi}=0$. In this case there are always two
tachyonic poles.
case(4) $\tilde{\kappa} \neq 0, m_{\phi} \neq 0$. This is the most general case. 
 Here there are always two poles. One pole is always tachyonic and the other
one behaves as in case(2). Let us define $\bar{s}$ as the solution to
$$1+\tilde{\kappa} a(\bar{s})=0.$$ Then the tachyonic pole is always located
to the left of $-\bar{s}$. The other pole can be tachyonic or physical depending
on the region of parameter space. This phenomena, where the pole switches from
unphysical to physical and vice-versa has been noted earlier in ref. \cite{chiv2}.
In fig.[11], we have depicted an example of how the pole positions change in this
manner as the parameters of the theory are varied.
Of course the subtraction procedure introduced earlier and the subsequent
renormalization must be carried out only for the unphysical poles. If $s_i$
denote the positions of the unphysical poles, then the corresponding
renormalization factors $z_i$ are given by:
\be
z_i^{-1}=1+{\Gamma \over [1+\tilde{\kappa}\bar{a}(s_i)]^2}\left({1 \over
s_i}-{2m_{\phi}^2 \over s_i^2 \bar{r}(s_i)}\ln{1+\bar{r}(s_i) \over
\bar{r}(s_i)-1}\right).
\label{multipole}
\ee
where, $$\bar{r}(s)= \sqrt{1+{4m_{\phi}^2 \over s}}$$ and $\bar{a}$ is the same
as $a$ with the replacement $r \rightarrow \bar{r}$.

\newpage

\begin{figure}[ht]
\centerline{\epsfig{file=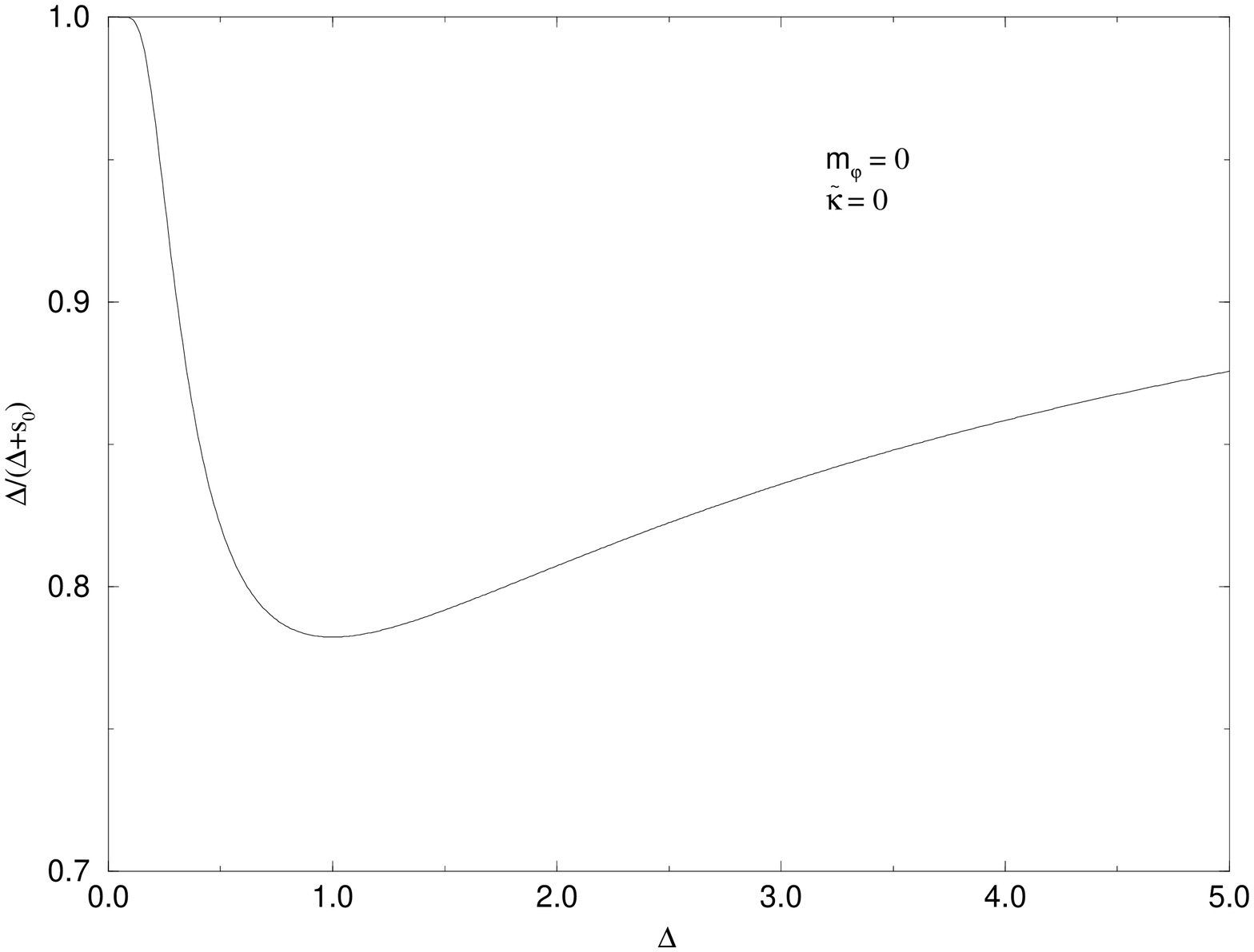,scale=0.7,angle=270}}
\caption{\label{fig1}
Non-perturbative propagator correction factor.}
\end{figure}

\begin{figure}[ht]
\centerline{\epsfig{file=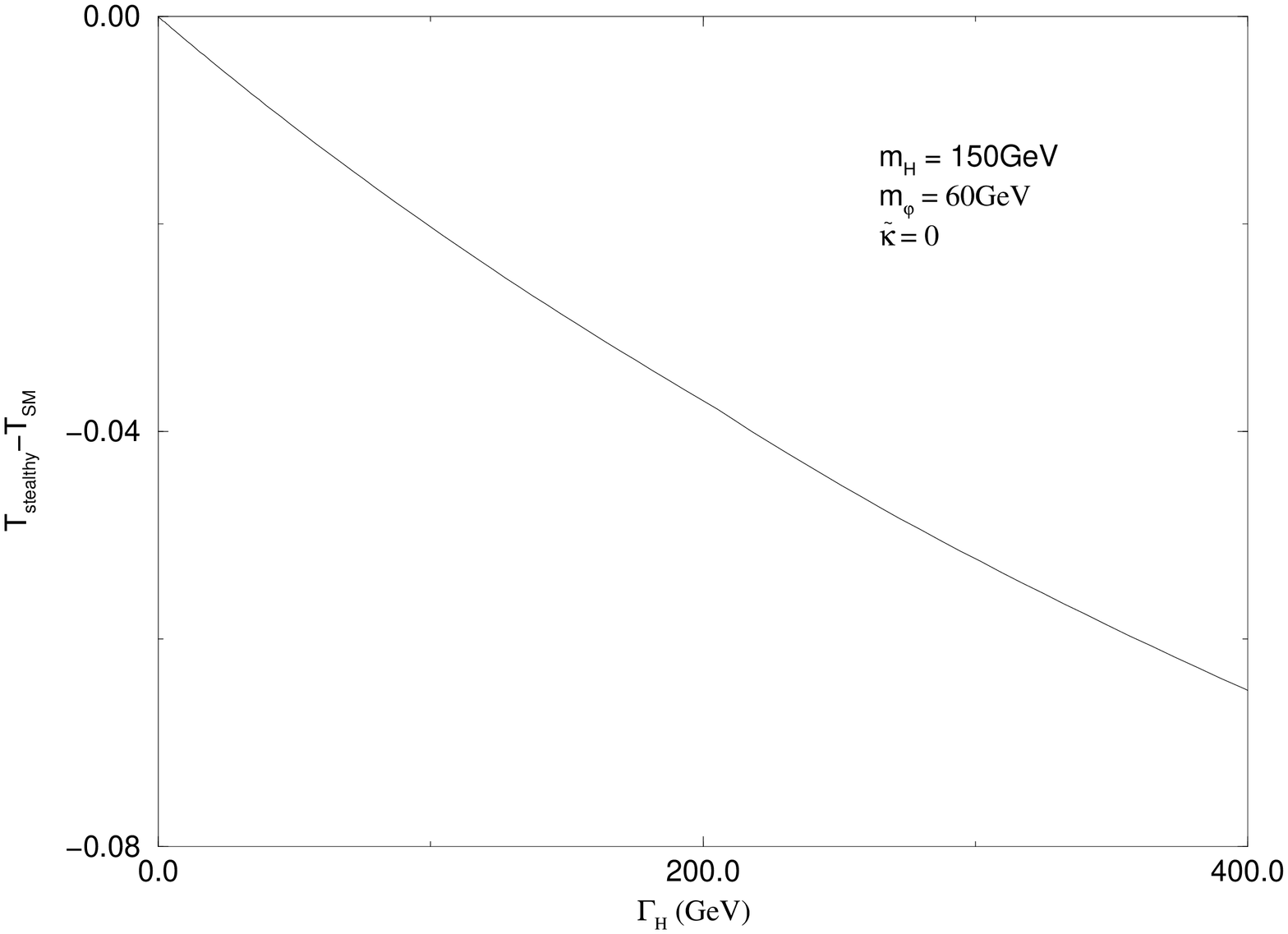,scale=0.7,angle=270}}
\caption{\label{fig2}
Correction to the T-parameter in the stealth model without self-interactions
among the phions.}
\end{figure}

\begin{figure}[ht]
\centerline{\epsfig{file=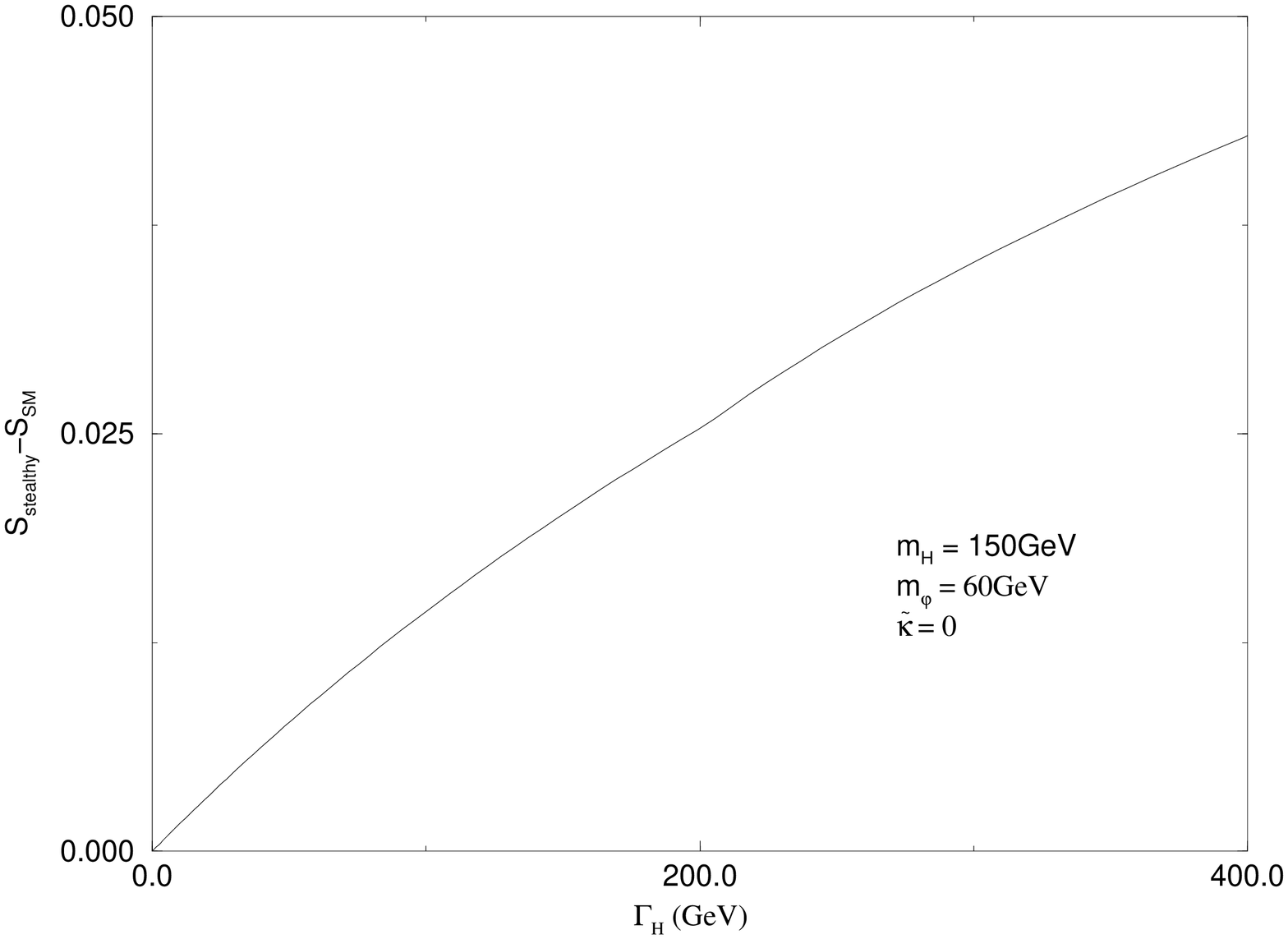,scale=0.7,angle=270}}
\caption{\label{fig3}
Correction to the S-parameter in the stealth model without self-interactions
among the phions.}
\end{figure}

\begin{figure}[ht]
\centerline{\epsfig{file=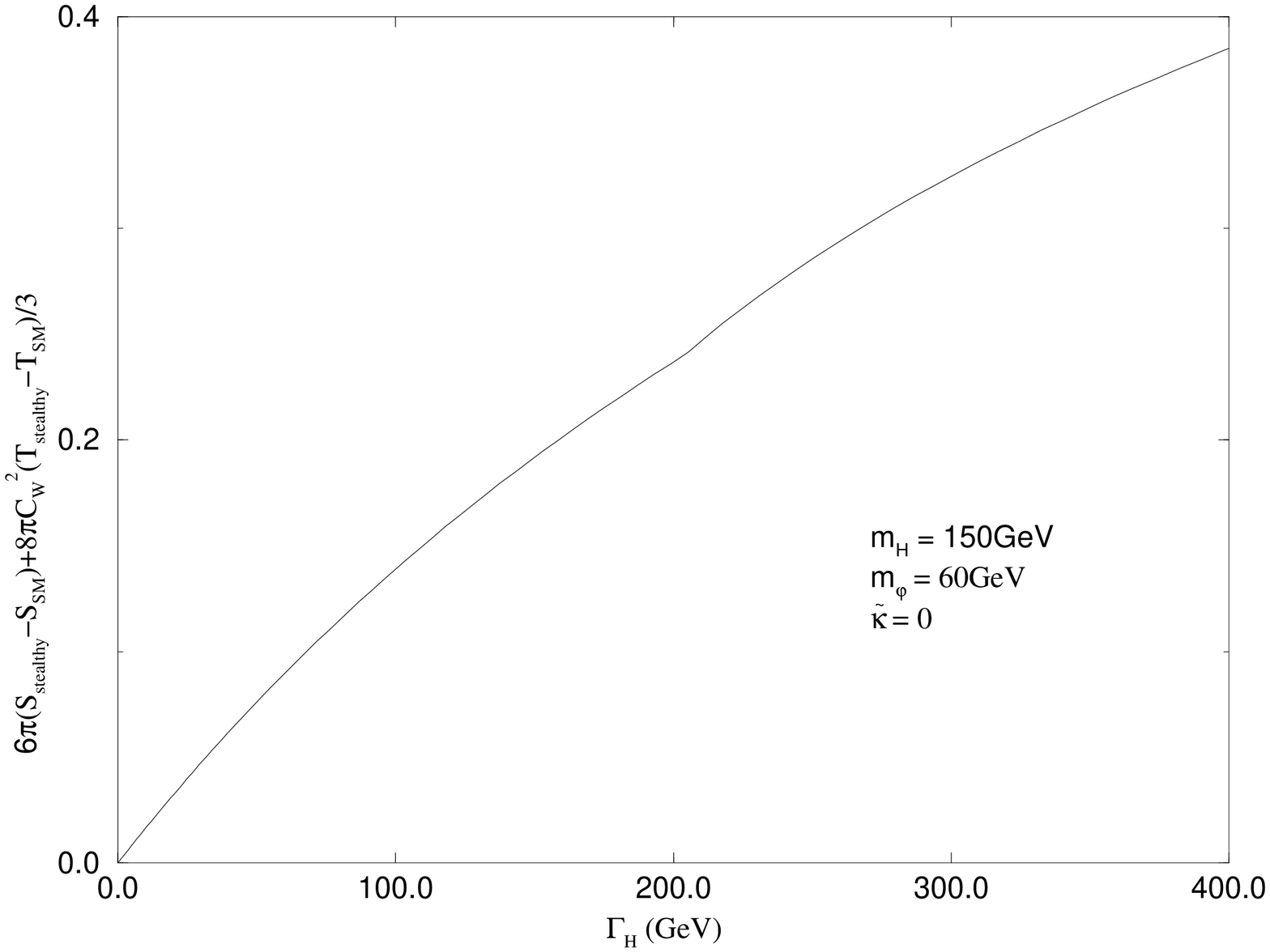,scale=0.7,angle=270}}
\caption{\label{fig4}
Correction to a (large) Higgs-mass independent quantity in the stealth model
without self-interactions among the phions.} 
\end{figure}

\begin{figure}[h]
\centerline{\epsfig{file=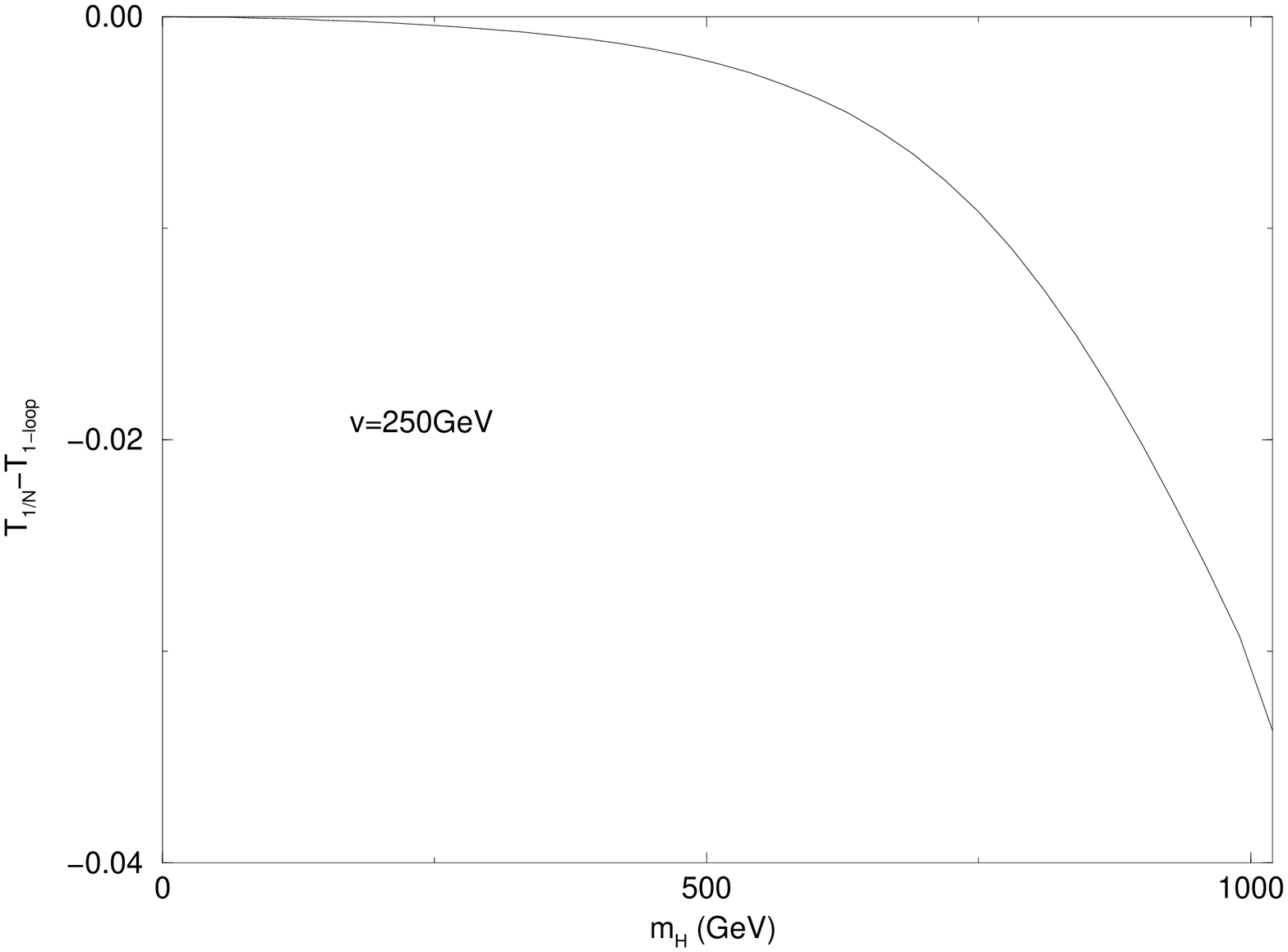,scale=0.7,angle=270}}
\caption{\label{fig5}
Correction to the T-parameter in the standard model, 1/N expansion minus
1-loop correction.}
\end{figure}

\begin{figure}[h]
\centerline{\epsfig{file=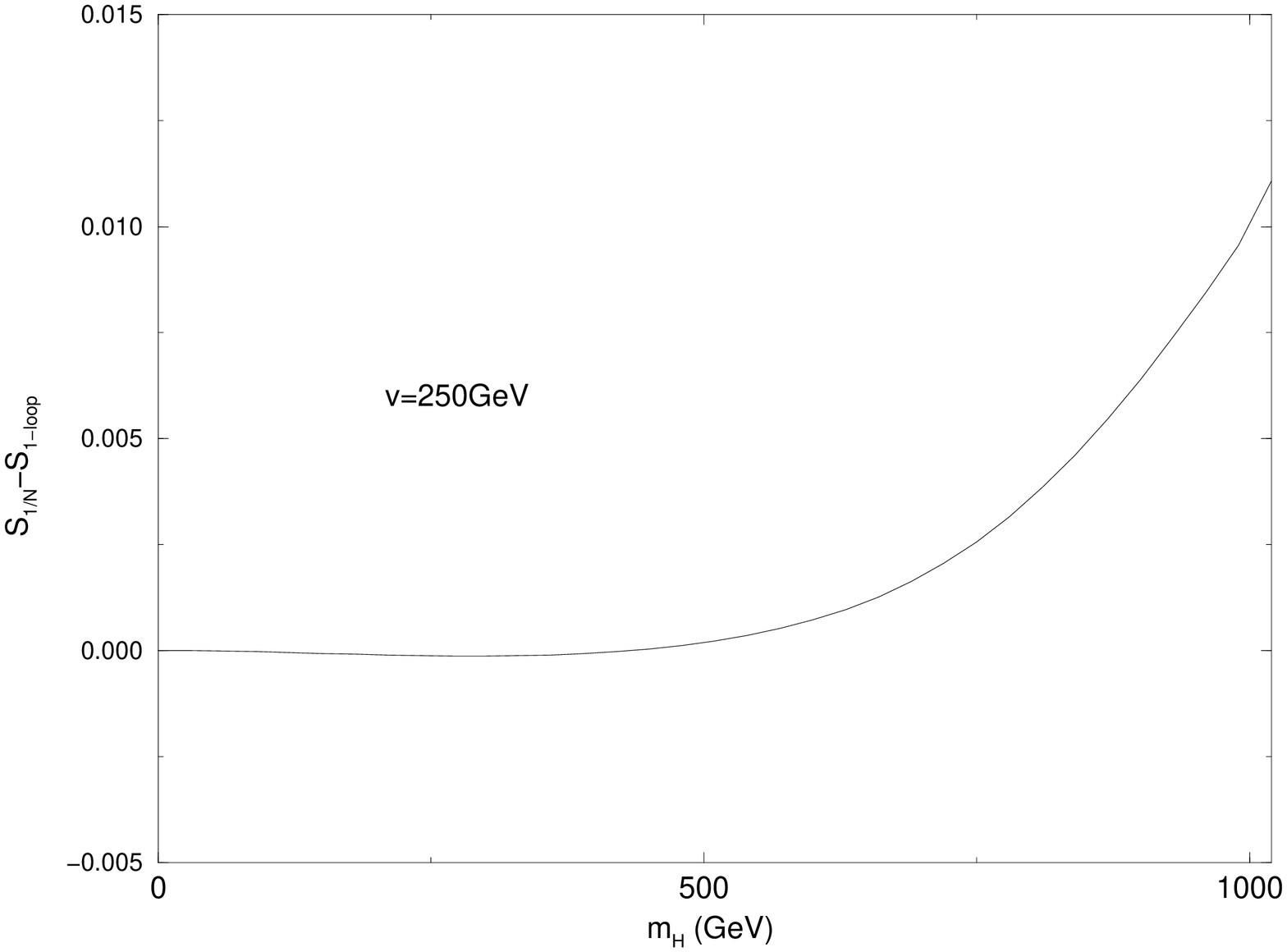,scale=0.7,angle=270}}
\caption{\label{fig6}
Correction to the S-parameter in the standard model, 1/N expansion minus
1-loop correction.}
\end{figure}

\begin{figure}[ht]
\centerline{\epsfig{file=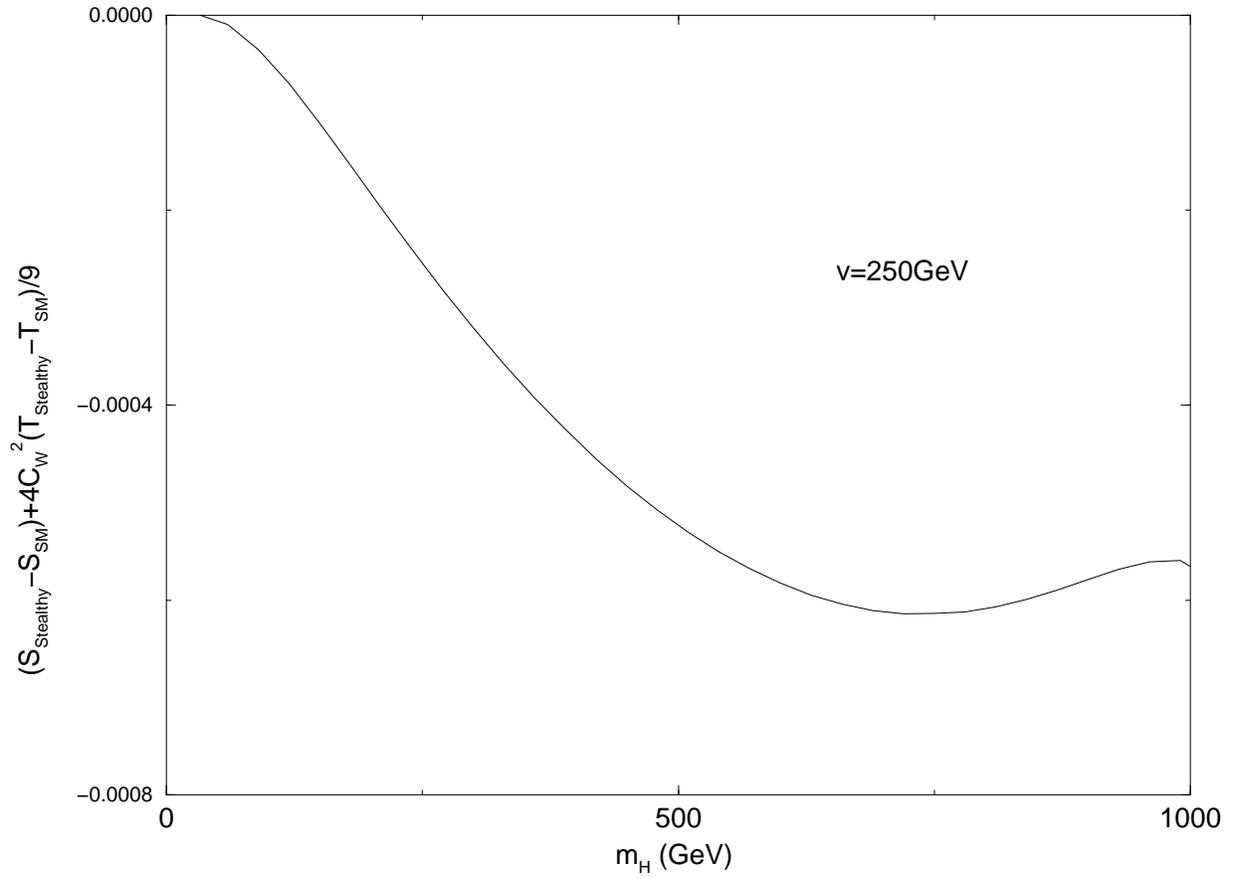,scale=0.7,angle=270}}
\caption{\label{fig7}
Correction to a 1-loop  (large) Higgs-mass independent quantity in the standard
mode, 1/N expansion minus 1-loop correction.} 
\end{figure}

\begin{figure}[ht]
\centerline{\epsfig{file=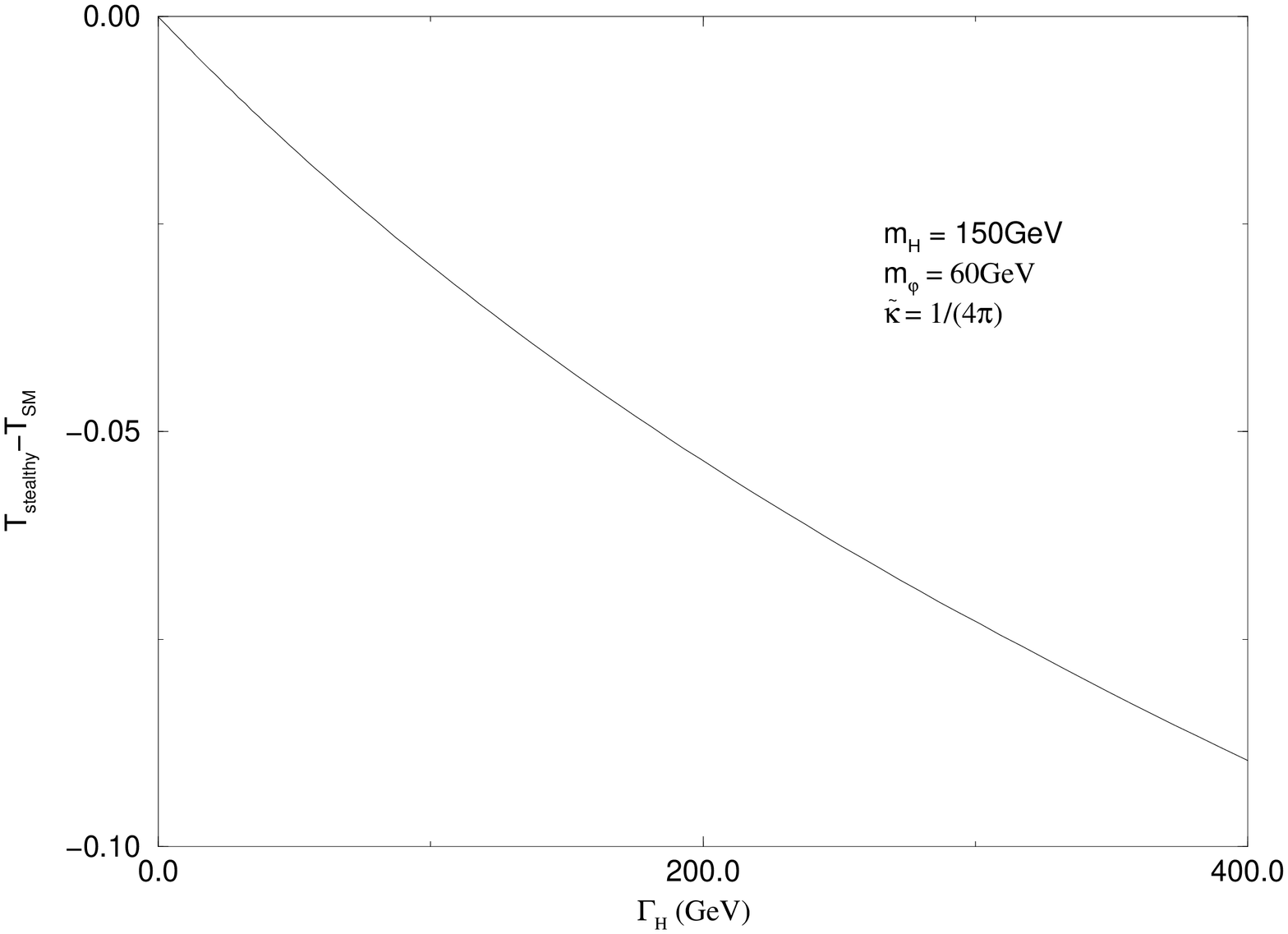,scale=0.7,angle=270}}
\caption{\label{fig8}
Correction to the T-parameter in the stealth model with self-interactions
among the phions.}
\end{figure}

\begin{figure}[ht]
\centerline{\epsfig{file=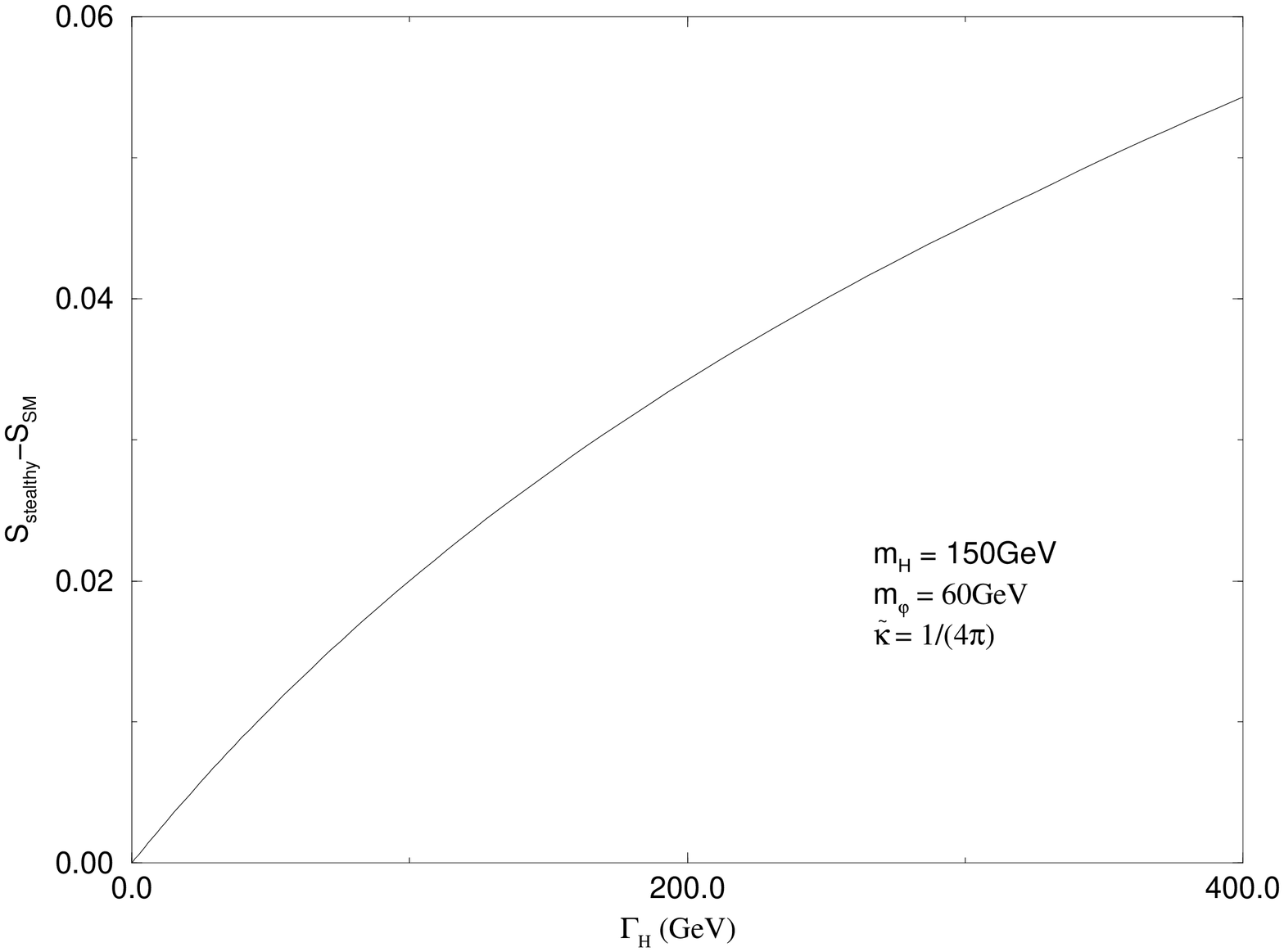,scale=0.7,angle=270}}
\caption{\label{fig9}
Correction to the S-parameter in the stealth model with self-interactions
among the phions.}
\end{figure}

\begin{figure}[ht]
\centerline{\epsfig{file=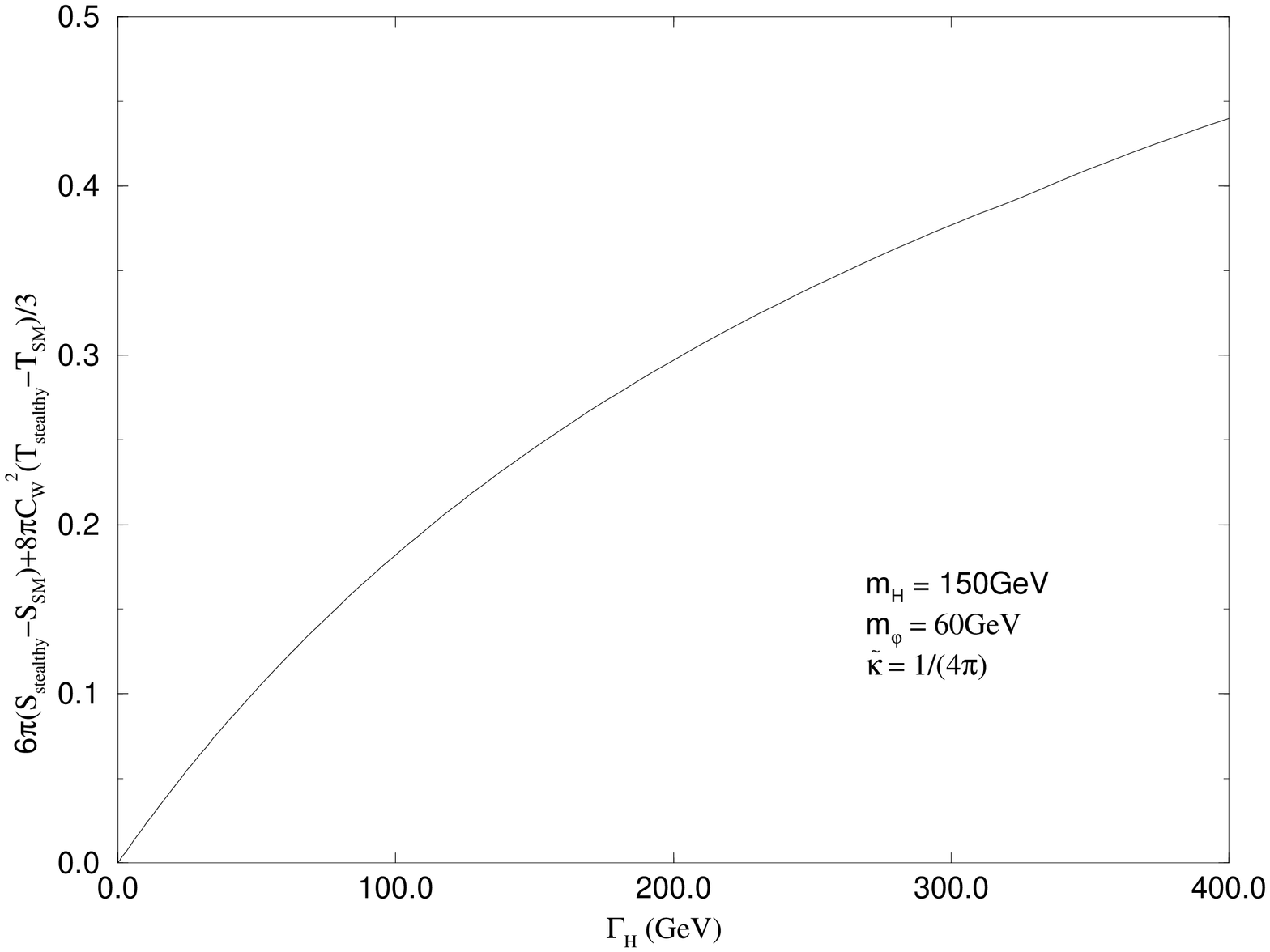,scale=0.7,angle=270}}
\caption{\label{fig10}
Correction to a (large) Higgs-mass independent quantity in the stealth model
with self-interactions among the phions.} 
\end{figure}

\begin{figure}[ht]
\centerline{\epsfig{file=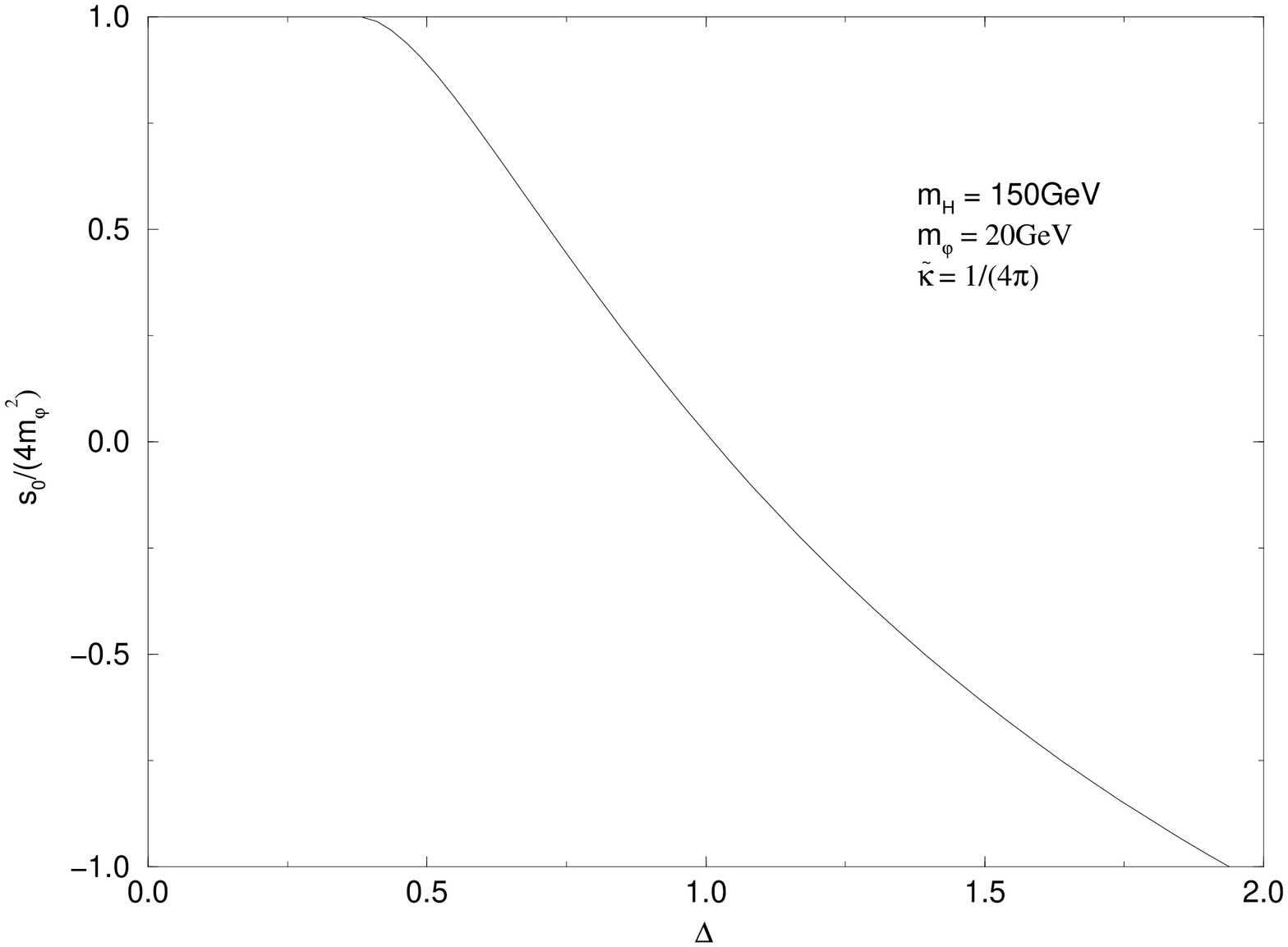,scale=0.7,angle=270}}
\caption{\label{fig11}
Change of pole position due to variation of the parameters.
}
\end{figure}


\begin{thebibliography}{99}
\bibitem{prec1}
P. Igo-Kemenes, ICHEP2000, Osaka 2000.

\bibitem{prec2}
A. Gurtu, ICHEP2000, Osaka 2000.

\bibitem{hill}
A. Hill, J. J. van der Bij, Phys. Rev. D36,3463 (1987).

\bibitem{gatto}
R. Casalbuoni, D. Dominici, R. Gatto, C. Giunti,
Phys. Lett. B178, 235 (1986).

\bibitem{chiv1}
R.S. Chivukula, M. Golden, Phys. Lett. B267, 233 (1991).

\bibitem{chiv2}
R.S. Chivukula, M. Golden, D. Kominis, M.V. Ramana,
Phys. Lett. B293, 400 (1992).

\bibitem{bjorken}
J.D. Bjorken, Int. J. Mod. Phys. A7, 4189 (1992). 

\bibitem{akh}
R. Akhoury, B. Haeri, Phys. Rev. D48, 1252 (1993).

\bibitem{binoth}
T. Binoth, J.J. van der Bij, Z. Phys. C75, 17 (1997).

\bibitem{krasnikov}
N.V. Krasnikov, Mod. Phys. Lett. A13, 893 (1998).

\bibitem{gunion}
J. F. Gunion, Phys. Rev. lett. 82, 1084 (1999).

\bibitem{ein1}
M. Einhorn, Nucl. Phys. B246, 75 (1984).

\bibitem{ein2}
M. Einhorn, H. Katsumata, Phys. lett. B181, 115 (1986).

\bibitem{cho}
P. Cho, Phys. Lett. B240, 407 (1990)

\bibitem{aoki}
K. Aoki, S. Peris, Z. phys. C61, 303 (1994).

\bibitem{bij1}
J.J. van der Bij, M.J.G. Veltman, Nucl.Phys. B205 (1984).

\bibitem{bij2}
J. J. van der Bij, Nucl. Phys. B248,141 (1984).

\bibitem{schwinger}
J. Schwinger, Phys. Rev. 82, 664 (1951)

\bibitem{olesen}
S. Chadha, P. Olesen, Phys. Lett. B72, 87 (1977);
P. Olesen, Phys. Lett. B73, 327 (1978).

\end{thebibliography}
\end{document}